\documentclass[aps,floats,amsmath,epsf,twocolumn, superscriptaddress, prl]{revtex4}
\usepackage{amsfonts, relsize, color}
\usepackage{graphics}
\usepackage{graphicx}
\usepackage{subfigure}
\usepackage{hyperref}
\usepackage{multirow}

\begin{document}
\title{Global Phase Diagram of a Three Dimensional Dirty Topological Superconductor}

\author{Bitan Roy}
\affiliation{Condensed Matter Theory Center and Joint Quantum Institute, University of Maryland, College Park, Maryland 20742-4111, USA}
\affiliation{Department of Physics and Astronomy, Rice University, Houston, Texas 77005, USA}

\author{Yahya Alavirad}
\affiliation{Condensed Matter Theory Center and Joint Quantum Institute, University of Maryland, College Park, Maryland 20742-4111, USA}

\author{Jay D. Sau}
\affiliation{Condensed Matter Theory Center and Joint Quantum Institute, University of Maryland, College Park, Maryland 20742-4111, USA}

\date{\today}

\begin{abstract}
We investigate the phase diagram of a three-dimensional, time-reversal symmetric topological superconductor in the presence of charge impurities and random $s$-wave pairing. Combining complimentary field theoretic and numerical methods, we show that the quantum phase transition between two topologically distinct paired states (or thermal insulators), described by thermal Dirac semimetal, remains unaffected in the presence of sufficiently weak generic randomness. At stronger disorder, however, these two phases are separated by an intervening thermal metallic phase of diffusive Majorana fermions. We show that across the insulator-insulator and metal-insulator transitions, normalized thermal conductance displays single parameter scaling, allowing us to numerically extract the critical exponents across them. The pertinence of our study in strong spin-orbit coupled, three-dimensional doped narrow gap semiconductors, such as Cu$_x$Bi$_2$Se$_3$, is discussed.   
\end{abstract}

\maketitle

\emph{Introduction}: Despite extensive band structure calculation over the second half of the last century, only in the past decade it became evident that seemingly boring band insulators can belong to two distinct families: (i) topological and (ii) trivial. Identifying insulators according to their topological nature culminated in a surge of theoretical and experimental investigations, leading to the successful realization of topological insulators in real materials~\cite{review-1, review-2}. Jurisdiction of topological classification, however, goes beyond insulators and encompasses semimetals (such as the Weyl semimetal) as well as superconductors (SCs), with our focus being on three dimensional, time-reversal symmetric paired states. Although candidates for such topological paired state are rather sparse, with the triplet $B$ phase of $^3$He standing as a prototypical example of charge-neutral topological superfluid~\cite{volovik}, a theoretical proposal alluding to the possibility of a \emph{charged} topological superconductor (TpSC) in strong spin-orbit coupled, doped narrow-gap (or gapless) semiconductors~\cite{fu-berg}, such as Cu$_x$Bi$_2$Se$_3$, Sn$_{1-x}$In$_{x}$Te, Nd$_x$Bi$_2$Se$_3$, Sr$_x$Bi$_2$Se$_3$, led to ample experimental investigations geared toward its possible material realization~\cite{exp-1, exp-2, exp-3, exp-4, exp-4a, exp-5, exp-6, exp-7, exp-8, exp-9, exp-10, exp-11}. However, the nature of superconducting order in these weakly correlated, but dirty materials still remains elusive, which, thus demands investigations on the role of randomness on topological pairing; constituting the central theme of the current Letter.

\begin{figure}[htb]
\includegraphics[width=8.75cm, height=3.65cm]{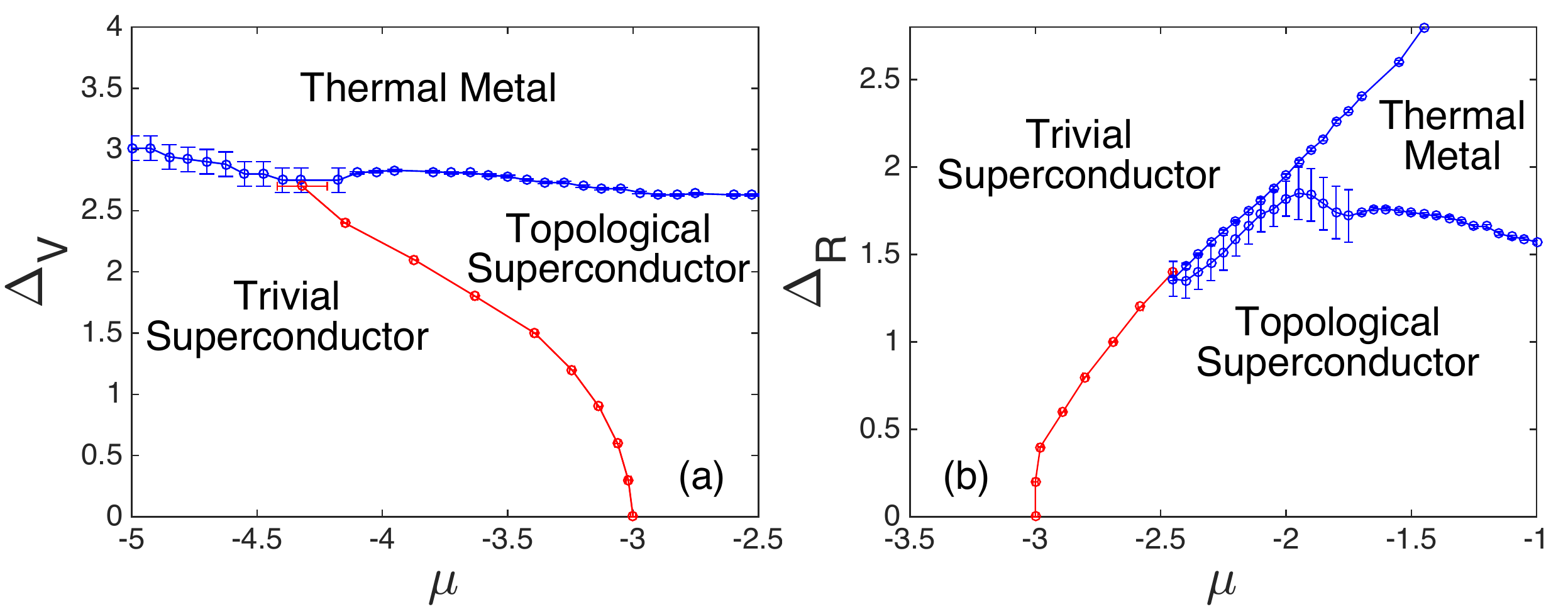}
\caption{Numerically obtained phase diagram of a three-dimensional class DIII system in the presence of random (a) charge impurities ($\Delta_V$), (b) $s$-wave pairing ($\Delta_R$). All transitions are continuous and the red lines represent a thermal Dirac semimetal. Various parameters from Eq.~(\ref{tightbinding}) are $t_1=2.5, t_2=1,n=1$ for (a), and $t_1=t_2=1$, $n=2$ for (b). } \label{pd-numerics}
\end{figure}

\begin{figure}[t!]
\includegraphics[width=8.75cm, height=3.65cm]{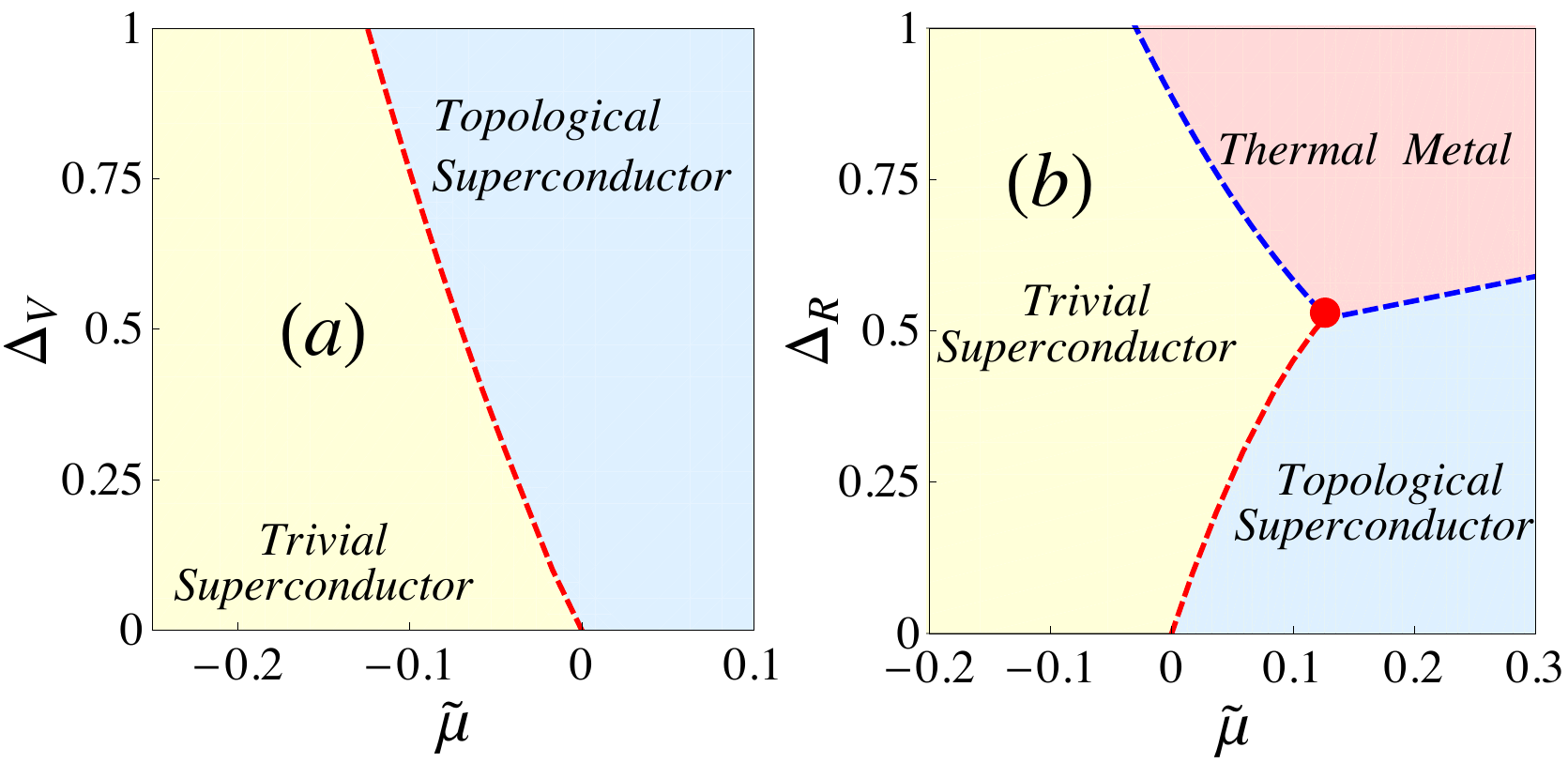}
\caption{ Phase diagram in (a) $\tilde{\mu}-\Delta_V$ and (b) $\tilde{\mu}-\Delta_R$ planes, obtained from leading order RG calculation [see Eq.~(\ref{RGEq})] for $b_0=0.5$. The massless Dirac fermion is realized along red lines, which undergoes a continuous QPT to a thermal metal through a multicritical point [red dot in (b)].}\label{pd-RG-1}
\end{figure}

Various three-dimensional doped narrow gap (or gapless) semicondcutors, such as Cu$_x$Bi$_2$Se$_3$, can support a fully gapped time-reversal symmetric, odd-parity topological superconductor (TpSC) at low temperatures~\cite{fu-berg, silaev}. Such TpSC belongs to class DIII and is characterized by an integer $Z$ invariant~\cite{classification}, manifesting through surface Andreev bound states that may also result in zero bias conductance peak (ZBCP)~\cite{exp-3, exp-4, exp-4a}. In addition, a three dimensional TpSC can give rise to a plethora of exotic phenomena, among which axionic electrodynamic~\cite{witten, roy-goswami-fieldtheory, sudbo}, $\theta$ vacuum for gravitational field~\cite{zhang, ryu}, cross-correlated responses~\cite{nagaosa}, propensity toward an axionic pairing~\cite{roy-goswami-axion} and one dimensional gapless modes inside the vortex core~\cite{silaev, roy-goswami-vortex}, are the fascinating ones. In the close proximity to the Fermi surface such TpSC also maps onto the B-phase of $^3$He~\cite{silaev}, and the resulting reduced BCS Hamiltonian takes the form of a massive Dirac Hamiltonian. The TpSC is realized through a weak coupling BCS paring of underlying normal femrions living in the vicinity of the Fermi surface (determined by the Debye frequency $\omega_D$), while the trivial pairing occurs in its absence (strong coupling pairing). The quantum phase transition (QPT) between a TpSC and a trivial superconductor (TrSC) takes place through a \emph{quantum critical point} (QCP), and massless Dirac fermion governs its universality class. Understanding the robustness of such QPT against the onslaught of randomness shapes the path of the current investigation, which we pave with complimentary field theoretic and numerical analyses by focusing on the emergent quadratic Hamiltonian inside paired phases. Our main achievements are (i) the universality class of the TpSC-TrSC QPT remains unaffected in the presence of sufficiently weak disorder, while (ii) a direct transition between TpSC and TrSC gets avoided by an intervening compressible thermal metallic phase of diffusive Majorana fermions (possessing finite lifetime and mean-free path) for stronger disorder, see Fig.~\ref{pd-numerics}. A nonlinear sigma model for a metal, see Refs.~\cite{senthil} and ~\cite{nomura} respectively for $d=2$ and $d=3$, is, however, inadequate to capture the universality class of the QPT between two thermal insulators for weak disorder, which is described by massless Dirac fermions or the onset of a metallic phase at stronger disorder that takes place through an instability of massless Dirac fermion across a multicritical point (see Fig.~\ref{pd-numerics} and ~\ref{pd-RG-1}), as the nonlinear sigma model is derived by integrating out Dirac fermions. Thus, we perform a renormalization group (RG) in the fermionic basis and obtain a good agreement between numerical findings and analytical predictions~\cite{criticality-explanation}.

\emph{Effective model}: The reduced BCS Hamiltonian, describing a class DIII system is $H=\sum_{\mathbf k} \Psi^\dagger_{\mathbf k} \hat{H}_{\mathbf k} \Psi_{\mathbf k}$. The four component spinor is defined as $\Psi^\dagger_{\mathbf k}=\left( c^\ast_{\uparrow, \mathbf k}, c^\ast_{\downarrow, \mathbf k}, c_{\downarrow, -\mathbf k}, -c_{\uparrow, -\mathbf k} \right)$, where $c^\ast_{s,{\mathbf k}}$, $c_{s,{\mathbf k}}$ are respectively quasiparticle creation and annihilation operators with momentum ${\mathbf k}$, and spin projection $s=\uparrow/\downarrow$, and 
\begin{eqnarray}\label{reducedBCS}
\hat{H}_{\mathbf k}= \left( \frac{\mathbf{k}^2}{2 m_\ast}-\tilde{\mu} \right) \gamma_0 + \left( \frac{\Delta_t}{k_F}\right) \: i \gamma_0 \gamma_j k_j.
\end{eqnarray}
Amplitude of the triplet $p$-wave pairing is $\Delta_t$, $k_F=\sqrt{2 m_\ast \tilde{\mu} }$ is the Fermi momentum, $m_\ast$ bears the dimension of mass and chemical potential $\tilde{\mu}$ is measured from the bottom of the conduction band. The $\gamma$ matrices are $\gamma_0= \tau_3 \otimes \sigma_0$, $\gamma_j=\tau_2 \otimes \sigma_j$ where $j=1,2,3$. Two sets of Pauli matrices $\boldsymbol \tau$ and $\boldsymbol \sigma$ respectively operate on Nambu and spin indices. Summation over repeated spatial indices is assumed and we set $\hbar=1$. The BCS Hamiltonian remains invariant under the reversal of time ($\mathcal T$), an emergent parity symmetry (${\mathcal P}$) defined as $\mathbf{k} \to -\mathbf{k}$ and $\Psi_{\mathbf{k}} \to \gamma_0 \Psi_{\mathbf{k}}$, but lacks the spin-rotational invariance. The TpSC occurs in the presence of a Fermi surface ($\tilde\mu>0$) and the band-inversion takes place at $|{\mathbf k}|=k_F$, while TrSC is realized for $\tilde{\mu}<0$ (no Fermi surface). The TpSC-TrSC QPT takes place at $\tilde{\mu}=0$. Notice for $\tilde\mu>0$, $\hat{H}_{\mathbf k}$ gives rise to a pseudo-spin Skyrmion texture in the momentum space and the Skyrmion number defines the integer topological invariant of class DIII~\cite{review-2, volovik}. Two fully gapped phases only support \emph{localized} quasiparticle excitation and stand as \emph{thermal insulators}. Across TpSC-TrSC QPT uniform Dirac mass ($\tilde{\mu}$) serves as the tuning parameter. The scaling dimension of Dirac mass $[\tilde{\mu}]=1$ fixes the correlation length exponent (CLE) $\nu=1$ at this QCP, which is, therefore, guaranteed to be stable against sufficiently weak disorder by the \emph{Harris criterion}, since $\nu=1>2/d$ for $d=3$~\cite{harris}.

\emph{Disorder}: All together this system is susceptible to \emph{four} types of elastic scatterers (without the constraint of ${\mathcal T}$). The corresponding Hamiltonian is
\begin{equation}\label{gen-disorder}
H_D= V_V (\mathbf x) \gamma_0 + V_R (\mathbf x) \gamma_5 + V_I (\mathbf x) i \gamma_0 \gamma_5 + V^{j}_M (\mathbf x) i \gamma_5 \gamma_j.
\end{equation}  
The first term represents random charge impurities, appearing as \emph{mass} disorder. The real (imaginary) component of random $s$-wave paring is accompanied by fermion bilinear $\Psi^\dagger \gamma_5 \left( i \gamma_0 \gamma_5 \right) \Psi$, assuming the form of \emph{axial} chemical potential (${\mathcal P}$, ${\mathcal T}$-odd \emph{pseudoscalar} mass). The last entry accounts for random magnetic impurities. Scaling analysis suggests that weak disorder is an \emph{irrelevant} perturbation for massless Dirac fermion~\cite{supplementary}. Therefore, the universality class of TpSC-TrSC QPT remains unchanged in the presence of infinitesimal randomness, as anticipated from the Harris criterion. Thus the topological invariant in class DIII continues to remain a well-defined quantity in the presence of weak disorder~\cite{classification}. The metallic phase, where disorder is a relevant perturbation, is topologically trivial. Recently, there has been a surge of analytical~\cite{fradkin, shindou-murakami, goswami-chakravarty, ominato-koshino, roy-dassarma, radzihovsky, nandkishore, altland, juricic} and numerical~\cite{imura, herbut, brouwer-1, brouwer-2, pixley-1, pixley-2, Ohtsuki, bera, hughes, pixley-rareregion} works, exploring the effects of disorder in regular Dirac and Weyl semimetals.

\begin{figure}
\includegraphics[width=8.75cm, height=3.65cm]{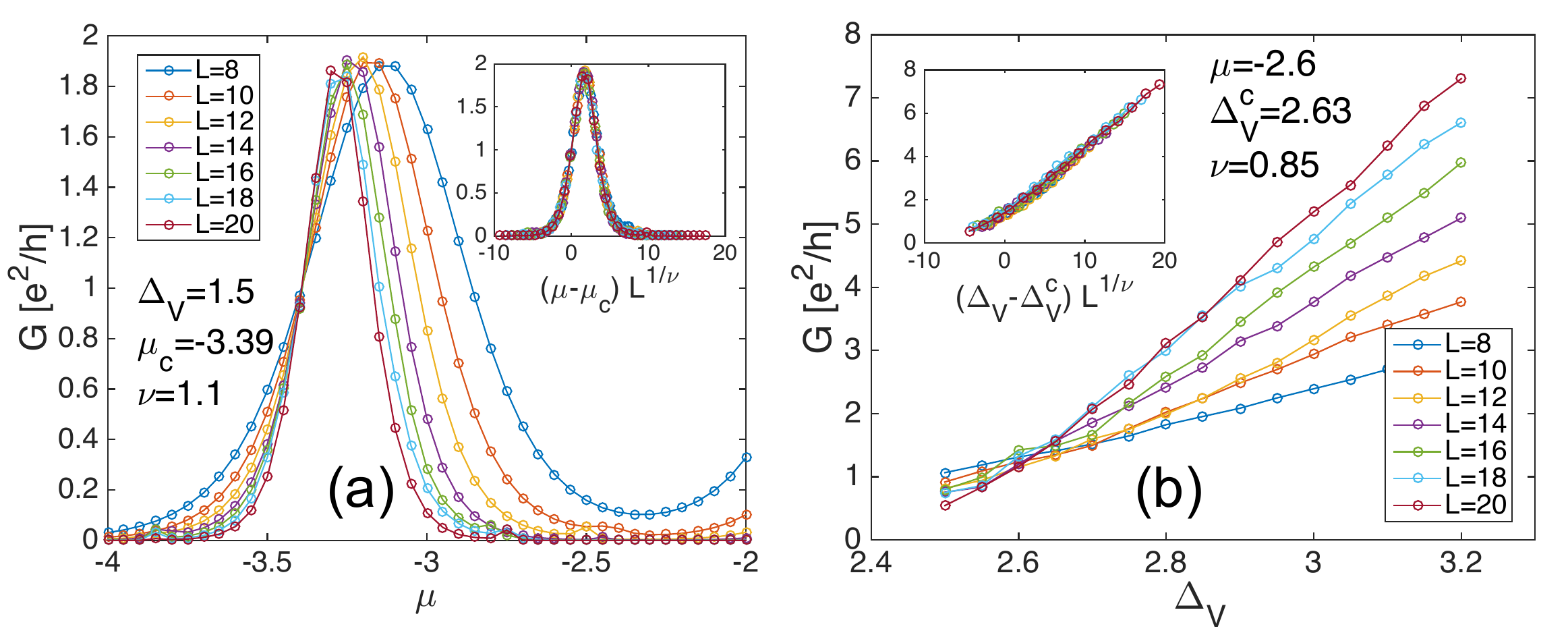}
\caption{Typical data collapse for the normalized thermal conductance (G) across (a) TpSC-TrSC and (b) SC-metal QPTs. The critical couplings and exponents are quoted in the figures. Parameters are same as in Fig.~\ref{pd-numerics}.  }\label{datacollapse}
\end{figure}

\emph{Numerical analysis}: The reduced BCS Hamiltonian from Eq.~(\ref{reducedBCS}) can be mimicked from a tight-binding model $H_L=\sum_{\mathbf k} \Psi^\dagger_{\mathbf k} \hat{H}^L_{\mathbf k} \Psi_{\mathbf k}$, which we implement on a cubic lattice of lattice spacing $a$ (set to be unity), where 
\begin{eqnarray}\label{tightbinding}
\hat{H}^L_{\mathbf k}=  i \gamma_0 \gamma_j t_1 \sin k_j- \gamma_0 t_2 \cos(n k_j) -\gamma_0 \mu + V_N (\mathbf x) \hat{N}.
\end{eqnarray}  
The effect of various disorder is captured by appropriate choice of $\hat{N}$ and $V_N (\mathbf x)$ [see Eq.~(\ref{gen-disorder})], which follows the \emph{standard normal distribution}. Typically we average over fifty disorder realizations. Without any loss of generality, for $\hat{N}=\gamma_0$ and $\gamma_5$, we respectively choose $n=1$ and $2$. The TpSC-TrSC QPT takes place at $\mu=-3 t_2$. To establish the phase diagram of this model, we numerically compute the normalized thermal conductance (NTC) $G=G_T/G_0$ (in units of $e^2/h$) in a cubic lattice with linear dimension $L$, where $G_0=(1/2) \times (\pi^2 k^2_B T_0)/(3 h)$ and $T_0$ is the temperature of the metallic reservoir~\cite{Wiedemann–Franz}. The factor of $1/2$ from particle-hole doubling in Eq.~(\ref{reducedBCS}). For numerical calculation of NTC, we use the Landauer formula and the software package KWANT~\cite{akhmerov}.

Gauge invariance mandates that electrical conductivity scales as $\sigma \sim L^{2-d}$. General scaling theory suggests that around a continuous phase transition the correlation length ($\xi$) diverges as $\xi \sim \delta^{-\nu}$, where $\delta$ is the reduced distance from the transition point, leading to the scaling ansatz $\sigma=L^{2-d} \mathcal{F}(L^{1/\nu} \delta)$, where $\mathcal{F}$ is an unknown but universal scaling function. Thus, NTC follows the scaling form $G=\mathcal{F}(L^{1/\nu} \delta)$. \emph{Respectively inside a thermal insulator and a metal $G$ and $G/L$ vanishes or saturates to a finite values in the thermodynamic limit ($L \to \infty$), while for massless Dirac fermion $G$ is independent of $L$.}

The phase diagram of a DIII system in the presence of random charge impurities is shown in Fig.~\ref{pd-numerics}(a). For weak disorder the TpSC-TrSC QPT takes place through band gap closing, where the system is described by massless Dirac fermion and NTC becomes $L$-independent. With increasing system size ($L$), the NTC monotonically decreases inside two thermal insulating phases, as shown in Fig.~\ref{datacollapse}(a). For a given strength of weak disorder all curves in the $G-\mu$ plane for various $L$, cross at a particular point, corresponding to the TpSC-TrSC transition point at $\mu=\mu_c$. An excellent data collapse is then obtained by comparing $G$ vs. $(\mu-\mu_c) L^{1/\nu}$ for $\nu=1.1 \pm 0.05$, as shown in Fig.~\ref{datacollapse}(a) (inset), in good agreement with the field theoretic prediction $\nu=1$. Following the same numerical approach we arrive at the phase diagram for a class DIII system in the presence of random $s$-wave (real) pairing, displayed in Fig.~\ref{pd-numerics}(b). The CLE across the insulator-insulator transition is $\nu=0.9 \pm 0.1$, also in good agreement with the field theoretic answer $\nu=1$~\cite{supplementary}. However, at stronger disorder a direct transition between two topologically distinct paired states gets avoided by an intermediate \emph{thermal metallic phase}, (see Fig.~\ref{pd-numerics}), suggesting that strong disorder is a relevant perturbation in this system. Next we substantiate these outcomes from a RG analysis.

\emph{RG analysis}: One-loop Wilsonian RG calculation leads to the following flow equations 
\allowdisplaybreaks[4]
\begin{eqnarray}\label{RGEq}
\frac{dv}{dl}=v\left[z-1-\sum_{N} \Delta_N \right], \: \: \frac{d \tilde{\mu} }{dl}=\tilde{\mu} +F\left(\Delta_N \right) b, \nonumber \\
 \frac{d \Delta_N}{d l}=-\Delta_N + \sum_{N,M} A_{N,M} \; \Delta_N \; \Delta_M, \: \: \frac{d b}{dl}=-b,
\end{eqnarray}     
in terms of the dimensionless variables $\Lambda/(2 m_\ast v)\to b$, $\tilde{\mu}/(v\Lambda) \to \tilde{\mu}$, $\Delta_N \Lambda/(2 \pi^2 v^2) \to \Delta_N$, after integrating out the fast Fourier modes within the momentum shell $\Lambda e^{-l}<|{\mathbf k}|<\Lambda$. Here $\Lambda \sim \frac{\omega_D}{v}$ is the ultraviolet cutoff, and $v=\frac{\Delta_t}{k_F}$ is the effective Fermi velocity of BdG quasiparticles. The coupling constant after disorder averaging over $V_N(\mathbf{x})$ is given by $\Delta_N$ for all $N$, $F(\Delta_N)=\Delta_V-\Delta_R-\Delta_I+3\Delta_M$ and $A_{N,M}$ is a $4 \times 4$ matrix~\cite{supplementary}. The scale invariance of $v$ leads to a scale dependent dynamic scaling exponent $z(l)=1+ \sum_N \Delta_N (l)$.

When disorder is weak $\Delta_N (l) \sim \Delta_{N,0} \; e^{-l}$, $b(l)\sim b_0 \; e^{-l}$ and $\tilde{\mu}(l) \sim \tilde{\mu}_0 \; e^{l}$. Quantities with the subscript ``$0$" denote their bare values. Therefore, the only relevant parameter in this regime is the Dirac mass ($\tilde{\mu}$), and we find 
\begin{eqnarray}\label{effectmass}
\tilde{\mu} (l)+\frac{b(l)}{3} \; F\left( \Delta_N (l) \right) \approx e^{l} \: \left[ \tilde{\mu}_0+\frac{b_0}{3}F\left( \Delta_{N,0} \right) \right].
\end{eqnarray}
The quantity on the left-hand side act as an effective Dirac mass, which when vanishes corresponds to the TpSC-TrSC QPT. For now we turn off $\mathcal{T}$-odd disorder.

 In the presence of random charge impurities, which is the dominant source of elastic scattering in any material, the phase boundary between TpSC and TrSC is determined by the condition $\tilde{\mu}=-b\Delta_V/3$ when $\Delta_V \ll 1$, as shown in Fig.~\ref{pd-RG-1}(a). By contrast, in the presence of random $s$-wave pairing the TpSc-TrSC QPT takes place when $\tilde{\mu}=b\Delta_R/3$ for $\Delta_R \ll 1$, as shown in Fig.~\ref{pd-RG-1}(b). Therefore, by increasing the strength of random charge impurities ($s$-wave pairing) one can drive the system from TrSC (TpSC) to TpSC (TrSC), but the reserve is not possible. These features at weak disorder are in qualitative agreement with numerical results; see Fig.~\ref{pd-numerics}.

The leading order flow equation of $\Delta_V$ does not support any nontrivial solution, naively suggesting absence of a metallic phase for strong charge impurities. The existence of a QPT for strong $\Delta_V$ and a resulting metallic phase can be established once we take into account higher order perturbative corrections~\cite{supplementary}. But, the appearance of a metallic phase through a MCP can be non-perturbative in nature, which therefore can only be accessed through exact numerical simulations. The flow equation of $\Delta_R$ suggests that there exists a MCP at $\Delta_R=\Delta^{c}_R=1/2$, and $\tilde{\mu}=b=0$ in the ($\tilde{\mu},\Delta_R$) plane [red dot in Fig.~\ref{pd-RG-1}(b)], where two thermal insulators and a compressible thermal metal, constituted by diffusive Majorana fermion meet, in qualitative agreement with numerical analysis. Beyond this point a direct transition between two topologically distinct thermal insulating phases get avoided by an intervening metallic phase, where the average density of states at zero energy is finite and the thermal conductivity $\kappa \sim T$ as $T \to 0$.

\begin{figure}[t!]
\includegraphics[width=8.75cm, height=3.65cm]{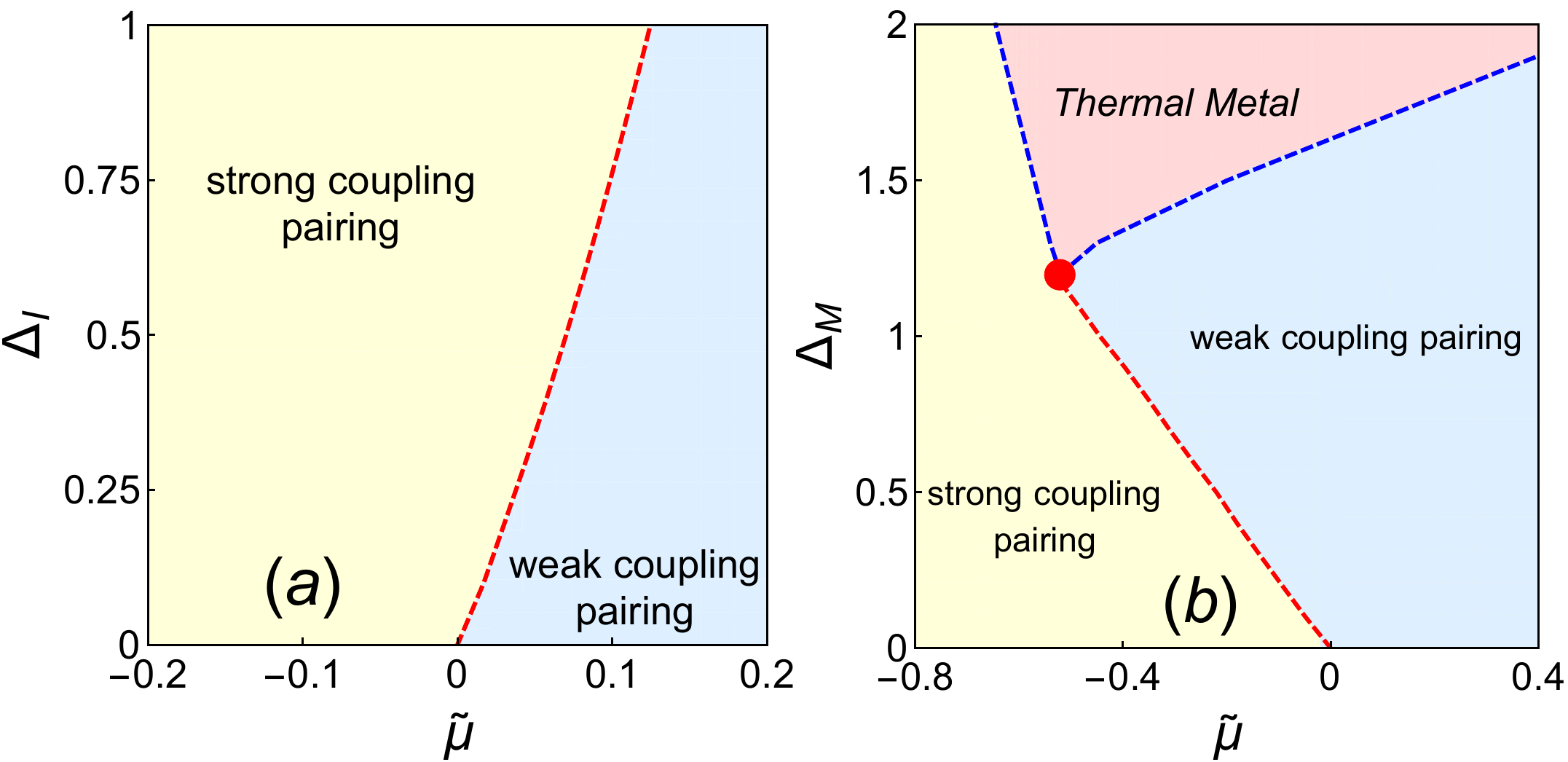}
\caption{Phase diagram in (a) $\tilde{\mu}-\Delta_I$, (b) $\tilde{\mu}-\Delta_M$ planes obtained from Eq.~(\ref{RGEq}) for $b_0=0.5$. Notations are same as in Fig.~\ref{pd-RG-1}. } \label{pd-RG-2}
\end{figure}

\emph{Metal-insulator transition (MIT)}: Our numerical analysis endows an opportunity to study the disorder induced MIT transition [see Fig.~\ref{pd-numerics}], across which NTC in different systems cross at a particular strength of disorder (for fixed $\mu$), signifying the transition point, as shown in Fig.~\ref{datacollapse}(b) for charge impurities. Comparing $G$ vs. $(\Delta_V-\Delta^c_V) L^{1/\nu}$ we obtain high quality data collapse for $\nu=0.85 \pm 0.05$ far from the MCP; see Fig.~\ref{datacollapse}(b) (inset). Across the MIT, driven by random $s$-wave pairing [see Fig.~\ref{pd-numerics}(b)], we find $\nu=0.7 \pm 0.1$ far from the MCP~\cite{supplementary}. Since quasiparticle excitation in the insulating (metallic) phase is localized (delocalized), as our numerical analysis on NTC would suggest, we believe that the correlation length ($\xi \sim |\delta|^{-\nu}$) represents the quasiparticle localization length (mean-free path), at least deep inside the phase.

\emph{$\mathcal T$-breaking disorder}: Now we briefly discuss the effects of $\mathcal T$-breaking disorder. Despite that in its presence the system enters into class $D$, which in three dimensions does not support any nontrivial topological invariant~\cite{classification}, one can still distinguish different pairing regimes. Based on scaling and RG analysis, QPT between thermal insulators is expected to remain unharmed even in the presence of $\mathcal T$-breaking disorder, see Fig.~\ref{pd-RG-2}. For sufficiently dense magnetic doping ($\Delta_M$), we expect the appearance of a metallic phase, as suggested by the leading order RG analysis, which also supports non-quantized \emph{thermal Hall conductivity} ($\kappa_{xy}$). A ${\mathcal T}$-breaking metallic phase may also appear when random $s$-wave (imaginary) pairing ($\Delta_I$) is strong enough, which, however, goes beyond the scope of leading order RG calculation~\cite{supplementary}.

\emph{Experiments}: Finally, we comment on the experimental relevance of our study. Early experiments showed that Cu$_x$Bi$_2$Se$_3$ becomes a superconductor around $T_c \sim 3$ K, however the nature of the pairing has remained controversial. While point contact spectroscopy only displays a fully gapped spectrum when $x \sim 0.2$~\cite{exp-5}, indicative of a trivial $s$-wave pairing, systems with $x\approx 0.3$ display ZBCP~\cite{exp-3,exp-4, exp-4a} (possible signature of a TpSC). Recent NMR and thermodynamic measurements are also suggestive of a triplet pairing for $x \sim 0.3$, with the $d$-vector locked in the $ab$-plane~\cite{exp-7, exp-9}, compatible with a \emph{triplet nematic} SC, which also represents a class DIII TpSC~\cite{liang-nematic-1, liang-nematic-2, supplementary}. \emph{Hence, it is quite reasonable to assume that with increasing Cu concentration ($x$), the strength of $s$-wave pairing decreases, while that of random charge impurity increases in the system.} With increasing $x$, nucleation of static $s$-wave pairing possibly ceases at a particular point $x=x_c$ (say), with $0.2 < x_c < 0.3$, and system can only sustain random $s$-wave pairing for $x>x_c$. Phase diagrams in Figs.~\ref{pd-numerics}, \ref{pd-RG-1}, and \ref{pd-RG-2}(a) strongly suggest that these effects can be conducive for direct BCS-BEC QPT, around which the specific heat $C_v \sim T^3$ and $\kappa \sim T^2$ (for thermal Dirac semimetal) or a MIT, with $C_v \sim T$ and $\kappa \sim T$ inside the metallic phase. Therefore, future experiments in (Cu/Nd/Sr)$_x$Bi$_2$Se$_3$, Sn$_{1-x}$In$_{x}$Te can unveil a rich phase diagram at low temperatures, with disorder playing the role of a \emph{nonthermal} tuning parameter.

\emph{Finite-T}: Although all the phases and phase transitions are strictly defined at $T=0$, our proposed global phase diagram can shed light on the scaling of various physical observables at finite-T at least when $T \ll T_c$. At finite temperature the above mentioned scaling of $C_v$ and $\kappa$ for thermal Dirac semimetal can be observed inside the quantum critical regime for $T>T_\ast = |X-X_c|^{\nu z}$ (but $T_\ast<T_c$) where $\nu z=1$ at the TpSC-TrSC QCP (for weak disorder) and $X=\mu$ or $\Delta$ is the tuning parameter. The critical scaling of physical observables at strong disorder is controlled by the MCP [see Figs.~\ref{pd-numerics}, ~\ref{pd-RG-1}]. Thus, at finite-T aforementioned scaling of $C_v$ and $\kappa$ inside the insulating or metallic phases can be observed when $T<T_\ast$. Since $T_c \sim 4$K in Cu/Nd/Sr$_x$Bi$_2$Se$_3$ and Sn$_{1-x}$In$_{x}$Te, distinct scaling behaviors of $C_v$ and $\kappa$ can be observed over a wide range of temperatures by varying the strength of impurities.

\emph{Conclusions}: To conclude, we establish that while the QPT between topological and trivial SCs remains unaffected in the presence of sufficiently weak randomness, a metallic phase of diffusive Majorana fermion masks the direct transition between them at stronger disorder. Our proposed global phase diagram of a dirty class DIII system (see Figs.~\ref{pd-numerics}, \ref{pd-RG-1}) can be relevant for strong spin-orbit coupled doped semiconductors~\cite{exp-1, exp-2, exp-3, exp-4, exp-4a, exp-5, exp-6, exp-7, exp-8, exp-9, exp-10, exp-11}, and $^3$He in aerogel~\cite{he3dirt-1,he3dirt-2}. It is worth mentioning that Fig.~\ref{pd-numerics}(a) also represents a phase diagram of class AII systems subject to mass disorder, where two gapped phases represent trivial and strong $Z_2$ topological insulators.

\emph{Acknowledgments}: This work was supported by JQI-NSF-PFC. B. R. is thankful to P. Goswami and D. F. Agterberg for discussions, and the Princeton Center for Theoretical Science for hospitality.

\emph{Note added}: After completing this work, we became aware of a preprint~\cite{Goswami-superuniversality}, also discussing the global phase diagram of class DIII system using the RG method.


\end{document}